\documentclass[10pt,twocolumn,letterpaper]{article}

\usepackage[pagenumbers]{cvpr} 

\definecolor{cvprblue}{rgb}{0.21,0.49,0.74}
\usepackage[pagebackref,breaklinks,colorlinks,allcolors=cvprblue]{hyperref}
\usepackage[most]{tcolorbox}
\usepackage[T1]{fontenc}

\usepackage{algorithm}
\usepackage{algpseudocode}
\def\paperID{3337} 
\def\confName{CVPR}
\def\confYear{2026}

\title{Adaptive Multi-Agent Reasoning for Text-to-Video Retrieval}
\author{Jiaxin Wu\textsuperscript{1,2}, Xiao-Yong Wei\textsuperscript{3,2}\thanks{Corresponding Author.}
,  Qing Li \textsuperscript{2}
\\
\textsuperscript{1}Shenzhen University, \textsuperscript{2}The Hong Kong Polytechnic University\\ \textsuperscript{3}Sichuan University, \\
{\tt\small jiaxin.wu@email.szu.edu.cn, cswei@scu.edu.cn,  qing-prof.li@polyu.edu.hk}\\
}

\begin{document}
\maketitle

\begin{abstract}
The rise of short-form video platforms and the emergence of multimodal large language models (MLLMs) have amplified the need for scalable, effective, zero-shot text-to-video retrieval systems. While recent advances in large-scale pretraining have improved zero-shot cross-modal alignment, existing methods still struggle with query-dependent temporal reasoning, limiting their effectiveness on complex queries involving temporal, logical, or causal relationships.
To address these limitations, we propose an adaptive multi-agent retrieval framework that dynamically orchestrates specialized agents over multiple reasoning iterations based on the demands of each query. The framework includes: (1) a retrieval agent for scalable retrieval over large video corpora, (2) a reasoning agent for zero-shot contextual temporal reasoning, and (3) a query reformulation agent for refining ambiguous queries and recovering performance for those that degrade over iterations. These agents are dynamically coordinated by an orchestration agent, which leverages intermediate feedback and reasoning outcomes to guide execution. We also introduce a novel communication mechanism that incorporates retrieval-performance memory and historical reasoning traces to improve coordination and decision-making.
Experiments on three TRECVid benchmarks spanning eight years show that our framework achieves a twofold improvement over CLIP4Clip and significantly outperforms state-of-the-art methods by a large margin. 
\end{abstract}

\section{Introduction}
\label{sec:intro}

With the exponential growth of short-form video platforms such as TikTok and YouTube Shorts, users are increasingly surrounded by massive volumes of dynamic, diverse, and unstructured video content \cite{grandview2024videoanalytics,pellicer2025shortvideo}. In this context, text-to-video retrieval, the task of retrieving relevant video segments from large-scale corpora using free-form natural language queries, has become essential for content discovery, organization, and navigation~\cite{wu2025genSearch, trecvid2023, zhangTextVideoRetrievalGLSCL2025}. Beyond user-facing applications, it has also emerged as a key augmentation module for multimodal large language models (LLMs), enabling them to retrieve visual evidence that supports multimodal reasoning and question answering~\cite{guo2025seed15vltechnicalreport, Qwen2.5-VL}.

\begin{figure}
  \centering
        \includegraphics[width=1\linewidth]{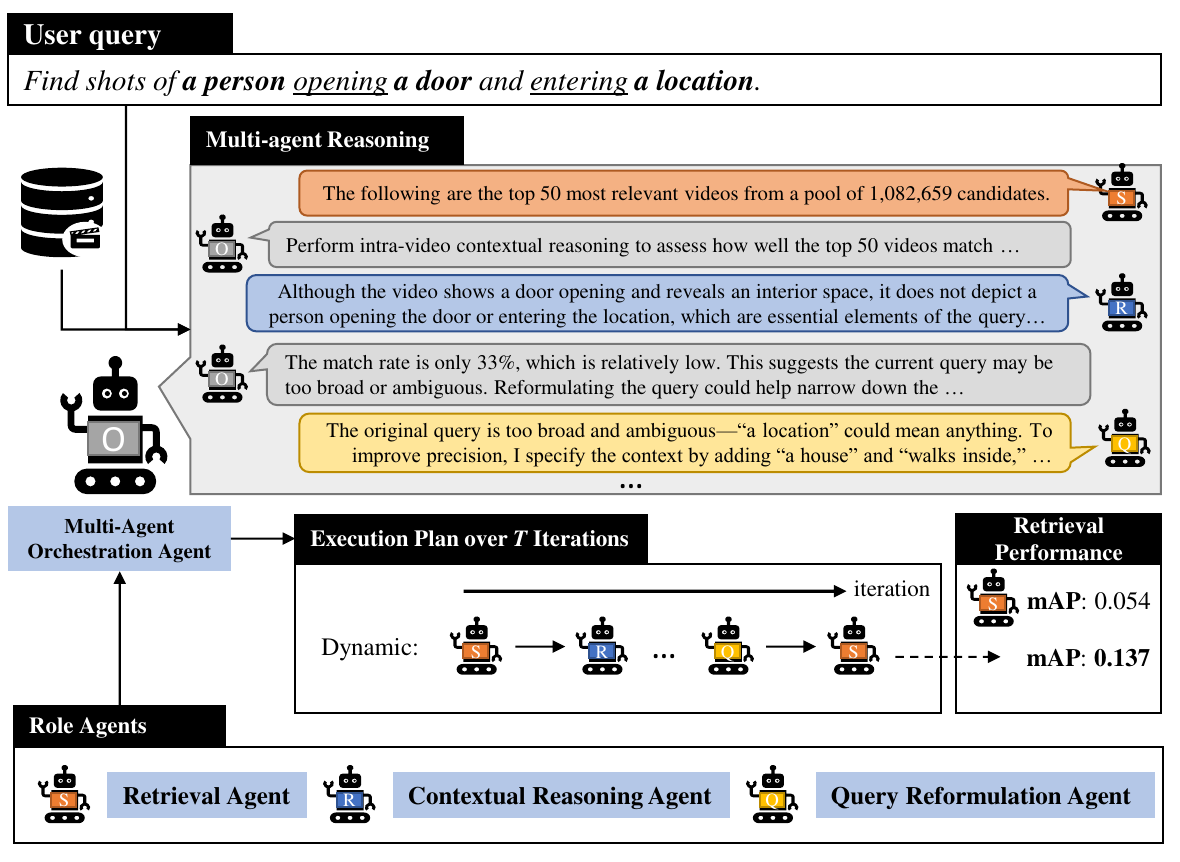}
      \caption{    Snapshot of our dynamic multi-agent retrieval framework applied to a complex query. The orchestration agent adaptively determines which role agents to invoke at each iteration based on intermediate feedback. Compared to retrieval-only baseline, the dynamic agent orchestration yields significantly higher retrieval performance.
}
\label{fig:snapshot}
    \vspace{-0.25in}

\end{figure}

Despite its importance and potential, text-to-video retrieval remains a highly challenging task \cite{trecvid2023,VBS2021,Jeong_2025_CVPR_audio-guided,Kumail2025CVPR_Negation}.
One major difficulty lies in the complex and diverse cross-modal alignment between free-form natural language queries and the temporally structured nature of video content. Specifically, queries may express a wide range of intentions, from identifying specific objects or actions to reasoning about temporal sequences, logical constraints (e.g., negation or conjunction), or causal relationships between events. As a result, effective retrieval requires contextual temporal reasoning, not only to recognize what appears in a video, but also to understand when and in what order events occur.
For example, consider the query: \textit{“Find shots of a person opening a door and entering a location”}. This requires identifying semantic entities (i.e., a person, a door, and a location), recognizing actions (i.e., opening and entering), and understanding their temporal relationship. Furthermore, it implies a logical co-occurrence constraint in which both actions must be performed by the same actor and occur within a continuous temporal segment.
A second challenge arises from the need to perform large-scale comparisons across millions of videos to identify relevant results while filtering out irrelevant ones. This goes beyond intra-video reasoning and necessitates scalable and efficient retrieval mechanisms.
Finally, in the era of multimodal LLMs, retrieval models are increasingly expected to operate in zero-shot settings, handling previously unseen queries and video content without task-specific fine-tuning. 
Together, the demands of intra-video temporal reasoning, large-scale cross-modal retrieval, and zero-shot generalization make text-to-video retrieval a complex and pressing research challenge.

Most recent video retrieval methods primarily focus on addressing the large-scale cross-modal retrieval challenge: identifying relevant videos while filtering out irrelevant ones in massive corpora~\cite{w2vv,hgr,dual_task,tmm2021-sea,zhangTextVideoRetrievalGLSCL2025}. These approaches typically learn a joint latent space through contrastive learning, where videos relevant to a given query are pulled closer together, and irrelevant ones are pushed farther apart.
Recent efforts in this direction have explored fine-grained cross-modal representations~\cite{RIVRL,hgr, dual_coding_improved}, and have incorporated additional modalities or cues to enhance joint space learning, such as audio~\cite{Jeong_2025_CVPR_audio-guided}, frame-level captions~\cite{Hur_2025_CVPR_narrative_video_retrieval}, and contrary concepts~\cite{wu2023unlikelihood}.
In parallel, to tackle the challenge of zero-shot generalization, recent advances in large-scale pretraining and text–image generation have inspired the development of effective zero-shot video retrieval models. These models leverage both existing paired datasets~\cite{Luo2021CLIP4Clip, zhangTextVideoRetrievalGLSCL2025} and synthetic data generated using vision–language models~\cite{wang2023internvid, improvedITV}.

However, existing methods still fall short in addressing  query-dependent temporal reasoning within a single video. While current models perform relatively well at identifying specific semantic entities or actions, they struggle with more complex forms of reasoning such as temporal, logical, and causal inference. For instance, Wang et al.~\cite{Understand_Negation_rumcc_MM} and Kumail et al.~\cite{Kumail2025CVPR_Negation} report that current retrieval models perform poorly on queries involving negation, and propose augmenting training with negation-specific examples to improve alignment. Wu et al.~\cite{wu2025genSearch} further show that existing methods fail on complex queries involving logical and spatial constraints, and introduce a novel query paraphrasing framework that reformulates such queries into images or simplified forms using generative models and LLMs. We also found that existing methods fail to reason about event temporal order. For example, for the previously mentioned query, CLIP4CLIP and state-of-the-art retrieval models (e.g., GenSearch \cite{wu2025genSearch}, GLSCL \cite{zhangTextVideoRetrievalGLSCL2025}) achieve a mean average precision of less than 0.05.

To address the fundamental challenges of complex video retrieval and the limitations of existing methods, we propose an adaptive multi-agent video retrieval framework.
As illustrated in Figure~\ref{fig:snapshot}, our approach dynamically assembles a query-specific agent workflow based on intermediate feedback and reasoning outcomes.
Unlike prior work that applies fixed pipelines, our framework adaptively invokes and coordinates specialized agents tailored to the unique reasoning demands of each query.
Specifically, inspired by recent advances in MLLMs for zero-shot video understanding~\cite{Qwen2.5-VL,ye2025mplugowl3}, we integrate a MLLM as a contextual reasoning agent that performs deep intra-video analysis on top-ranked candidate videos, enabling fine-grained temporal and semantic interpretation. However, applying MLLMs directly over large-scale corpora is computationally infeasible. 
To address this, we introduce a scalable retrieval agent based on a text-to-video embedding model that efficiently ranks videos in a joint space. This allows the reasoning agent to focus only on the most promising candidates.
Besides, to handle complex or ambiguous queries, our framework includes a query reformulation agent, which decomposes queries into simpler, visually grounded sub-queries. This enhances retrieval success by aligning the query structure with the retrieval model’s capabilities. 
More importantly, these agents do not operate in isolation. Their collaboration is coordinated by a central LLM-based controller, referred to as the multi-agent orchestration agent. This controller monitors performance at each step and dynamically decides whether to continue exploiting current results or initiate exploration via reformulation.

Beyond the core multi-agent architecture, we design a communication mechanism that facilitates coordination among agents and ensures a consistent reasoning process. In particular, to enhance the effectiveness of query reformulation, the reformulation agent must be aware of the retrieval agent’s capabilities. To support this, we introduce a retrieval-performance memory bank that stores previous search queries along with their evaluated performance, as assessed by the reasoning agent. This memory serves as a reference for the query reformulation agent, allowing it to adapt its strategies to the retrieval agent’s strengths and limitations. 
In addition, we share the orchestration agent’s decisions and reasoning traces with subsequent agents to ensure consistency across execution steps.

We validate our method on three TRECVid Ad-hoc Video Search benchmarks. Experimental results demonstrate that it doubles the retrieval performance of recent pretraining model \cite{zhangTextVideoRetrievalGLSCL2025}, and outperforms recent text-to-video retrieval systems on over 72\%  queries \cite{wu2025genSearch}. Our proposed communication mechanism further contributes to significant improvements over the baseline. Beyond performance gains, the framework also provides interpretable reasoning traces, offering valuable insights into the system's decision-making and query reformulation processes.

\begin{figure*}
    \centering
        \includegraphics[width=1\linewidth]{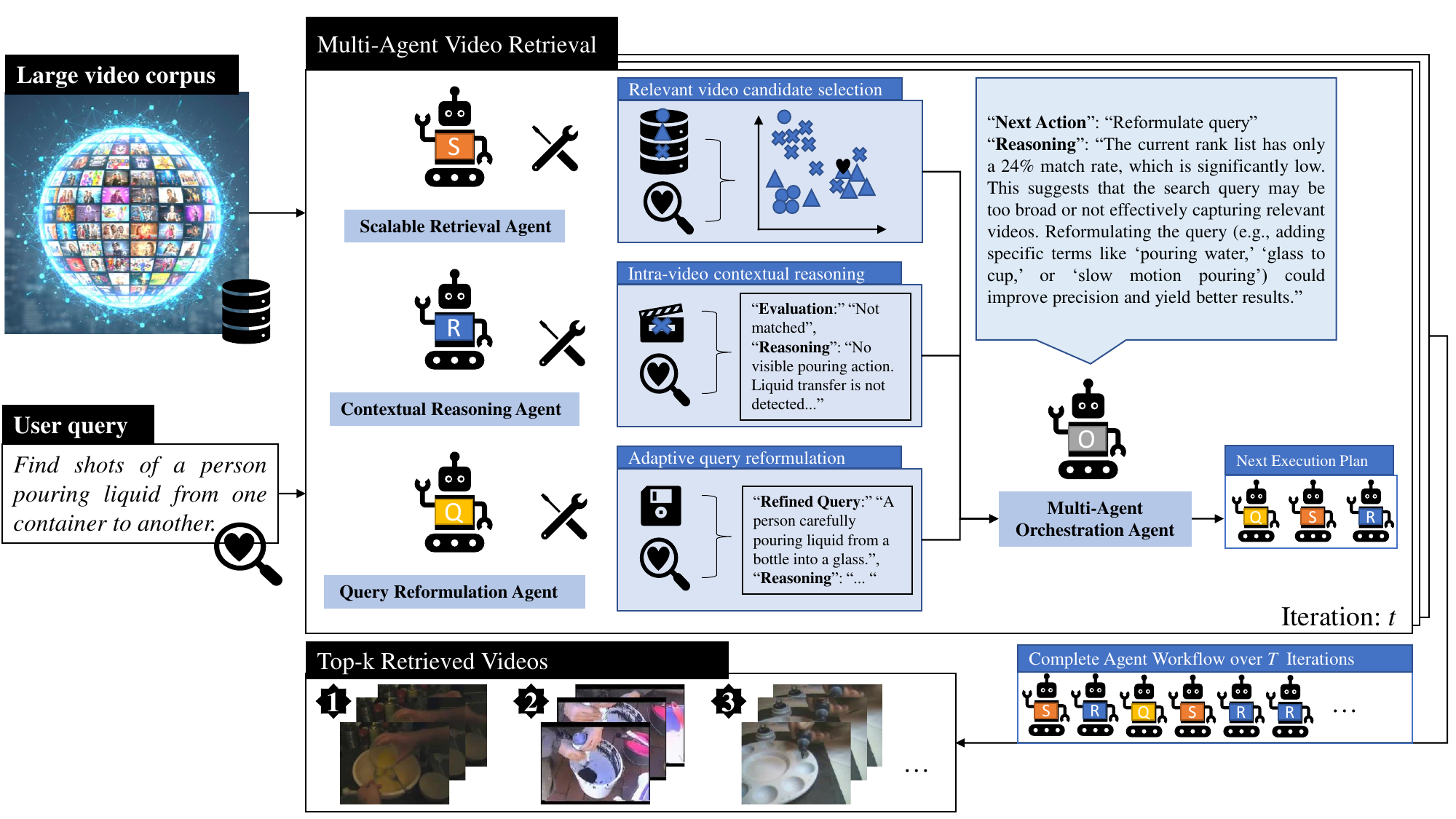}
    \caption{Overview of our multi-agent video retrieval framework. Given a user query, a scalable retrieval agent (S) selects top-ranked candidate videos from a large corpus. A contextual reasoning agent (R) conducts fine-grained intra-video analysis of the retrieved results, while a query reformulation agent (Q) adapts the query in response to ambiguous or low-quality matches. A multi-agent orchestration agent (O) dynamically determines the execution plan, deciding which agents to invoke at each iteration, based on intermediate reasoning signals and past retrieval outcomes. }
    \label{fig:framwork}
\end{figure*}



\section{Related Work}

\subsection{Text-to-Video Retrieval}
The history of open-vocabulary text-to-video retrieval dates back to 2006, when TRECVid introduced a pilot ad-hoc video search evaluation based on retrieval systems~\cite{trecvid2006}. Existing methods mainly focus on addressing cross-modal alignment between queries and large-scale video corpora. Early approaches extract key concepts from queries and videos and align them based on a predefined set of semantic categories~\cite{Waseda2016,Informedia2016}, while more recent methods leverage cross-modal embedding spaces to model semantic similarity between video segments and natural language queries~\cite{w2vv,dualconding}.
Recent advancements in large-scale pretrained vision-language models (VLMs) (e.g., CLIP~\cite{CLIP}, InternVid~\cite{wang2023internvid}, and BLIP-2~\cite{BLIP2}), along with more sophisticated architectures for modeling fine-grained semantic relationships~\cite{hgr} and representations~\cite{dual_task,RIVRL,improvedITV,tmm2021-sea,Jeong_2025_CVPR_audio-guided}, have further improved embedding-based approaches. However, these methods still fall short when handling complex queries involving fine-grained visual attributes, temporal dependencies, logical constraints, or causal relationships~\cite{Hur_2025_CVPR_narrative_video_retrieval, wu2025genSearch}. 
To address challenges in logical reasoning, recent work has introduced negation-aware learning~\cite{Kumail2025CVPR_Negation,Understand_Negation_rumcc_MM} and LLM-based query paraphrasing frameworks~\cite{wu2025genSearch}. However, instead of targeting only specific types of queries (e.g., negation), our work aims to develop a generalizable framework that can dynamically adapt to a wide range of complex queries by leveraging multi-agent collaboration and contextual temporal reasoning.

\subsection{Agent-based Video Retrieval Methods}

To the best of our knowledge, we are the first to introduce an agent-based approach for automatic text-to-video retrieval. While some prior work in interactive video retrieval has explored the concept of multi-agent systems composed of components, such as query classification, refinement, and result verification, with a programmatic controller managing their sequential workflow~\cite{agent_visual_intereactive}. These efforts primarily focus on image retrieval. Moreover, they lack detailed methodological illustration and empirical validation.
In contrast, we propose a multi-agent video retrieval framework specifically designed to address the challenges of contextual and temporal reasoning in text-to-video retrieval. Our approach incorporates dynamic coordination between agents and adaptively balances exploration and exploitation through an orchestrated agent workflow. Furthermore, we systematically evaluate and analyze the contribution of each agent to overall performance improvements.

\section{Multi-Agent Retrieval Framework}

Given a free-form query $q$ and a large video corpus $\mathcal{\textbf{C}} = \{v_i\}_{i=1}^N$ containing $N$ videos, the task of text-to-video retrieval is to return a ranked list of relevant videos using a retrieval model, i.e., $\hat{y} = f(\mathcal{\textbf{C}} \mid q)$, where $\hat{y} = [\hat{v}_1, \hat{v}_2, \dots, \hat{v}_k]$ denotes the top-$k$ videos ranked by their relevance to the query. The objective is to rank ground-truth videos relevant to $q$ higher than irrelevant ones.

In this paper, we propose an adaptive multi-agent retrieval framework to generate an effective ranked list, as shown in Figure~\ref{fig:framwork}. We reformulate the retrieval process as a sequence of $T$ reasoning iterations, where elements of the ranked list are incrementally collected through the collaborative interaction of specialized agents. Formally, the final output is $\hat{y} = [\hat{y}_1, \dots, \hat{y}_T]$, where each $\hat{y}_t$ represents a sub-ranked list retrieved at iteration $t$.
Specifically, our framework consists of four key agents:
1)\textbf{ A scalable retrieval agent }($f_S$) that performs efficient candidate selection from the corpus;
2)\textbf{ A contextual reasoning agent} ($f_R$), powered by a multimodal large language model (MLLM), that performs zero-shot reasoning over candidate videos, incorporating temporal, logical, and causal context;
3)\textbf{ A query reformulation agent }($f_Q$) that refines or decomposes the original query $q_0$ to improve alignment with visual content, particularly for complex or ambiguous queries; 
and (4) \textbf{a multi-agent orchestration agent} ($f_O$), powered by an LLM, that adaptively picks agents based on intermediate reasoning outcomes to balance exploration and exploitation over iterations.
In the following subsections, we detail how these agents operate and collaborate to produce a final submission rank list $\hat{y}$.

\begin{algorithm}[t]
\caption{\textit{Multi-agent Retrieval}}
\label{alg:MAR}
\textbf{Require:} video collection $\mathcal{\textbf{C}}$, original query $q_0$, retrieval agent $f_S$, reasoning agent $f_R$, query reformulation agent $f_Q$, and orchestration agent $f_O$, max iterations $T$, examination length $k$, submission length $L$.

\begin{algorithmic}[1]
\Statex \Comment{\textbf{Initialization Phase}}

\State Initialize submission set $\hat{\textbf{y}} \gets [\ ]$
\State Initialize memory bank $\mathcal{M} \gets \emptyset$
\State Initialize remaining video set $\hat{\mathcal{C}} \gets \mathcal{\textbf{C}}$
\State Initialize examination window $e \gets [0, k]$
\State $r_0 \gets f_S(q_0, \hat{\mathcal{C}})$

\Statex \Comment{\textbf{Iterative Decision and Evaluation Loop}}
\For{$t = 1$ to $T$}

    \State \Comment{Evaluate previous retrieval}
    \State $\hat{y}_{t-1}^{\text{match}}, \hat{y}_{t-1}^{\text{unmatch}} \gets f_R(p_{\text{eval}}, q_0, r_{t-1}[:k])$
    \State $h_{t-1}^{\text{eval}} \gets (k, |\hat{y}_{t-1}^{\text{match}}|, |\hat{y}_{t-1}^{\text{unmatch}}|)$
    \State $p_{t-1} \gets |\hat{y}_{t-1}^{\text{match}}| / k$
    \State \textsc{UpdateSearchSpace}$(\hat{\mathcal{C}}, r_{t-1}[:k])$
    \State \textsc{UpdateMemory}$(\mathcal{M}, t-1, q_{t-1}, p_{t-1}, e[0], e[1])$

    \State \Comment{Update submission set}
    \State \textsc{AppendSubmission}$(\hat{y}, \hat{y}_{t-1}^{\text{match}})$

    \If{$|\hat{y}| \geq L$}
        \State \textbf{break}
    \EndIf

    \State \Comment{Decide next action}
    \State $a_t, a_t^r \gets f_O(p_{\text{action}}, h_{t-1}^{\text{eval}})$

    \If{$a_t = \text{``explore''}$}
        \State $q_t, q_t^{r} \gets \hat{f}_Q(\hat{p}_{\text{refine}}, q_0, q_{t-1}, \mathcal{M},a_t^r)$
        \State $e \gets [0, k]$
    \ElsIf{$a_t = \text{``exploit''}$}
        \State $q_t \gets q_{t-1}$
        \State $e \gets [e[0] + k, e[1] + k]$
    \EndIf

    \State \Comment{Retrieve next ranked list}
    \State $r_t \gets f_S(q_t, \hat{\mathcal{C}})$

\EndFor

\State \Return $\hat{y}$ \Comment{Final submission set}
\end{algorithmic}
\end{algorithm}

\subsection{Scalable Retrieval Agent}

In real-world scenarios, text-to-video retrieval typically operates on a large corpus, often containing over a million videos~\cite{Trecvid2019,VBS2021}. To address scalability, we first apply a scalable retrieval agent to the entire corpus to select a subset of candidate videos that are semantically similar to the query. These candidates are passed to the reasoning agent for fine-grained evaluation. The ranked list $r_t$ at iteration $t$ is obtained as:
\begin{equation}
    r_t = f_S(q_t, \textbf{C}_t)
\end{equation}
Here, $f_S$ is a cross-modal retrieval model that computes similarity scores between the current query $q_t$ and the remaining video set $\textbf{C}_t$ ($\hat{\mathcal{C}})$, which excludes all videos previously examined by $f_R$ in iterations $1$ through $t{-}1$.

\subsection{Contextual Reasoning Agent}

We then apply contextual reasoning to the top-$k$ videos in $r_t$ using an MLLM and a reasoning prompt $p_{\text{eval}}$. The reasoning agent evaluates each candidate for semantic, temporal, logical, and causal alignment with the query. The output consists of two disjoint sets:

\begin{equation}
\hat{y}^{\text{match}}_t, \hat{y}^{\text{unmatch}}_t = f_R( p_{\text{eval}}, q_t, r_t[:k])
\end{equation}
A video in the match list $\hat{y}^{\text{match}}_t$ is considered relevant if it satisfies intra-video contextual reasoning (e.g., satisfying temporal order, causal relationships, or logical constraints) and is added to the final ranked list: $\hat{y}_t = \hat{y}^{\text{match}}_t$. Videos in $\hat{y}^{\text{unmatch}}_t$ are excluded from further consideration.

\subsection{Query Reformulation Agent}

In cases where the retrieval agent fails to retrieve relevant videos, particularly for complex queries involving negation, logical constraints, or compositional reasoning, we incorporate a query reformulation agent within the multi-agent framework. The reformulation agent takes the original query $q_0$ and the query used in the previous iteration $q_{t-1}$ as input, and generates a refined query $q_t$:

\begin{equation}
\label{eq:reformulation_agent_init}
q_t = f_Q( p_{\text{refine}}, q_0, q_{t-1})
\end{equation}
Here, $f_Q$ denotes the query reformulation function implemented with a large language model (LLM), guided by a prompt $p_{\text{refine}}$. 

\subsection{Multi-agent Orchestration Agent}

In the multi-agent framework, the reasoning agent ($f_R$) is responsible for exploitation (analyzing and filtering candidate videos), while the query reformulation agent ($f_Q$) facilitates exploration by generating alternative queries. During the iterative retrieval process, determining when to activate exploration is crucial for maintaining a high retrieval rate of relevant videos.

To address this, we introduce an orchestration agent that determines whether to trigger query reformulation at each iteration. It takes as input the evaluation history $h^{\text{eval}}_{t-1}$ from the previous step and returns both an action $a_t$ and a textual explanation $a_t^r$:
\begin{equation}
\label{eq:decision_agent_init}
a_t, a_t^r = f_O(p_{\text{action}}, h^{\text{eval}}_{t-1})
\end{equation}
Here, $f_O$ is an LLM-based decision function guided by a prompt $p_{\text{action}}$. The action $a_t \in \{\texttt{explore}, \texttt{exploit}\}$ determines whether the system should continue with the current query or initiate a reformulation step.

\subsection{Multi-agent Communication Mechanism}

Communication between agents is essential for effective collaboration. For instance, the query reformulation agent must understand the retrieval agent’s strengths and limitations to adapt its strategy accordingly.
To facilitate this coordination,  we introduce a communication mechanism in the form of a retrieval-performance memory bank $\mathcal{M} = [(t, q_t, h^{\text{eval}}_t, e^{\text{start}}, e^{\text{end}})]$, which stores the historical performance of past queries along with the corresponding examination windows $e = [e^{\text{start}}, e^{\text{end}}]$ of the ranked lists. The query reformulation function is then extended from $f_Q$ to $\hat{f}_Q$ as follows:
\begin{equation}
\label{eq:reformulation_agent_new}
q_t, q_t^r = \hat{f}_Q(\hat{p}_{\text{refine}}, q, q_{t-1}, \mathcal{M}, a_t^r)
\end{equation}
Here, $q_t^r$ denotes the reasoning behind the reformulated query, providing interpretability and traceability of the reformulation process. The reformulation function also incorporates the action reasoning $a_t^r$, ensuring consistency between decisions and query updates.
The complete retrieval process is described in Algorithm~\ref{alg:MAR}.


\section{Experiments}

\begin{table*}[t]
\caption{Performance comparison on TRECVid AVS datasets. The number in parentheses denotes the number of queries in each query set. } 
\label{tab:comparison_state-of-art}
\centering
\resizebox{\linewidth}{!}{
\begin{tabular}{l|ccc|ccc|cc|c}
    \toprule
Datasets                     & \multicolumn{3}{c|}{IACC.3}               & \multicolumn{3}{c|}{V3C1}   & \multicolumn{2}{c|}{V3C2}   \\
\hline
Query sets                         & tv16 (30)         & tv17 (30)           & tv18  (30)          & tv19 (30)   &tv20 (20) &tv21 (20) &tv22 (30) &tv23 (20)  &Mean     \\
                          \hline
\multicolumn{5}{l}{TRECVid top results:}                                                  \\
Rank 1     & 0.054  & 0.206  & 0.121 & 0.163   &0.359   &0.355   &\textbf{0.282} &0.292 &\\
\hline

W2VV ~\cite{w2vv}                      & 0.149          & 0.198          & 0.103          & /  & /   & /   & /    & /    & /   \\  
Dual-task ~\cite{dual_task}              & 0.185          & 0.241           & 0.123        &0.185    & /   & /    & /           & / & /\\ 

Dual coding ~\cite{dualconding}              & 0.160          & 0.232           & 0.120         & 0.163  & 0.208    & /     & /    & /    & /  \\ 

HGR ~\cite{hgr}             &/         & /           & /       & 0.142   &0.301     & /    & /           & /    & /     \\ 
CLIP  \cite{CLIP}       &   0.182 & 0.217    & 0.089  &0.117   & 0.128  &  0.178  &0.124 & 0.109 &0.143 \\
CLIP4CLIP \cite{Luo2021CLIP4Clip} & 0.182 &0.217 &0.089 &0.133 & 0.149 &0.188 & 0.121 & 0.109 &0.149\\
BLIP-2 \cite{BLIP2}     &0.213 &0.226 & 0.168&0.199 &0.222 &0.273 &0.164 & 0.203 &0.209 \\

LAFF \cite{LAFF}             &    0.188   &   0.261        &  0.152   &  0.215     &  0.299   &  0.300  & 0.178 &0.172  &0.221\\

RIVRL~\cite{RIVRL}  & 0.159          & 0.231           & 0.131        & 0.197    & 0.278  & 0.254 & 0.179  & 0.177 &0.201 \\
ITV \cite{wu2023unlikelihood}                &  0.211    & 0.292   &   0.170  &   0.227          &0.345  & 0.318 & 0.150  & 0.170 &0.235 \\
InternVid \cite{wang2023internvid} & 0.185 & 0.216 &0.104 &0.122 &0.199 &0.190 &0.125 &0.103 & 0.155\\
LanguageBind  \cite{zhu2024languagebind} & 0.182&0.238&0.145& 0.162 &0.221 &0.254 &0.184&0.146 &0.192\\
 IITV \cite{improvedITV} & 0.280 &	0.349 &	0.165 &	0.242	&0.352	&0.365 &	0.235&	0.295 &0.285\\
GLSCL \cite{zhangTextVideoRetrievalGLSCL2025}  &0.132 &0.185 & 0.075 &0.142 & 0.206 & 0.182& 0.127&0.119 &0.146\\
GenSearch  \cite{wu2025genSearch}       & 0.281	&0.366 &0.180	&0.263	&0.359	&0.377	&0.258	&0.311 &0.299\\
\hline
                   
Multi-agent Retrieval &\textbf{0.298} &\textbf{0.371} & \textbf{0.219}&\textbf{0.283}&\textbf{0.379}& \textbf{0.420}& 0.271 &\textbf{0.342}& \textbf{0.323} \\

\bottomrule
\end{tabular}
}
\end{table*}

\subsection{Datasets and Evaluation Metric}

We validate the proposed multi-agent retrieval framework on three TRECVid Ad-hoc Video Search (AVS) benchmarks: IACC.3~\cite{Trecvid2016}, V3C1~\cite{Trecvid2019}, and V3C2~\cite{trecvid2022}. These video corpora are collected from open video-sharing platforms, such as YouTube and Vimeo.
The IACC.3 dataset contains about 334,000 videos, while V3C1 and V3C2 are significantly larger, comprising about 1 million and 1.4 million video clips, respectively. The duration of video clips ranges from 1 to 202 seconds.
Each year, TRECVid releases 20–30 new ad-hoc video search queries targeting one of these large-scale corpora. We report retrieval performance on eight years of query sets for which ground-truth annotations are available. In total, 210 queries have been released. Each query typically expresses semantic constraints involving objects, people, actions, and locations, and may also include logical conditions (e.g., co-occurrence, negation), temporal ordering, causal relationships, or combinations thereof.

Following the TRECVid AVS setting \cite{trecvid2006,Trecvid2016,Trecvid2019,trecvid2022}, we report the inferred average precision (xinfAP), an estimation of mean average precision, as the evaluation metric. The evaluation/submission length $L$ is set to 1,000, consistent with the TRECVid and other approaches.

\subsection{Implementation Details}

We adopt IITV~\cite{improvedITV}, an interpretable cross-modal retrieval model, as our retrieval agent. For the large language model (LLM) and vision-language reasoning modules, including the orchestration, reasoning, and reformulation agents, we use Qwen3-VL-8B-Instruct~\cite{Qwen2.5-VL} which is capable of multimodal data understanding and reasoning as the backbone. 

All experiments are conducted on a server equipped with NVIDIA RTX A6000 GPUs (46GB VRAM). The whole framework can be run in one A6000 GPU. We use PyTorch as the deep learning framework, and Qwen3-VL is accessed via the HuggingFace Transformers library.
The maximum retrieval iteration is set to $T = 60$, and the examination window length is set to $k = 50$. 
We design prompt templates for each agent module: contextual reasoning ($p_{\text{eval}}$), query reformulation ($p_{\text{refine}}$ or $\hat{p}_{\text{refine}}$), and decision-making ($p_{\text{action}}$ or $\hat{p}_{\text{action}}$). These prompts are provided in the Appendix.
We also conduct a sensitivity analysis on the hyperparameters $T$ and $k$, and report the corresponding results in ablation study.

\subsection{Comparison with State-of-the-Art Methods}

We compare our proposed method against a broad range of state-of-the-art approaches, including top-performing systems reported by TRECVid and recent advances in video retrieval.
Specifically, W2VV is one of the earliest embedding-based methods to surpass traditional concept-based approaches on the AVS task. Building upon this foundation, methods such as Dual Coding~\cite{dualconding}, HGR~\cite{hgr},  and LAFF~\cite{LAFF} introduce more fine-grained visual-textual representations and architectural enhancements for improved retrieval performance.
Interpretable embedding-based methods include Dual-task~\cite{dual_task}, ITV~\cite{wu2023unlikelihood}, and RIVRL~\cite{RIVRL}, which aim to align vision and language modalities in a more transparent and explainable manner.
We also compare with recent large-scale pretraining approaches such as CLIP~\cite{CLIP}, CLIP4Clip~\cite{Luo2021CLIP4Clip}, BLIP-2~\cite{BLIP2}, InternVid~\cite{wang2023internvid}, and LanguageBind~\cite{zhu2024languagebind}, which have demonstrated strong performance on general cross-modal retrieval tasks.
Finally, we include comparisons with methods specifically designed to handle complex video queries, including IITV~\cite{improvedITV}, GenSearch~\cite{wu2025genSearch}, and the recent cross-modal pre-trained model GLSCL~\cite{zhangTextVideoRetrievalGLSCL2025}.

Table~\ref{tab:comparison_state-of-art} compares the retrieval performance of these approaches across three TRECVid AVS datasets. 
Our multi-agent retrieval method significantly outperforms all compared methods across every query set. 
Specifically, the mean performance of our method over the eight-year evaluation period exceeds that of the most recent pre-trained model (i.e., GLSCL) by a factor of 2.212.
Besides, compared with recent approaches that also target complex queries, i.e., IITV~\cite{improvedITV} and GenSearch~\cite{wu2025genSearch}, our multi-agent method achieves higher performance on 82.9\% and 74.3\% of the queries, respectively.

\begin{table*}[t]
\caption{The performance comparison of query reformation agent with and without memory bank} 
\label{tab:query_reformulation_comparison}
\centering
\resizebox{\linewidth}{!}{
\begin{tabular}{l|ccc|ccc|cc|c}
    \toprule
Datasets                     & \multicolumn{3}{c|}{IACC.3}               & \multicolumn{3}{c|}{V3C1}   & \multicolumn{2}{c|}{V3C2}   \\
\hline
Query sets                         & tv16 (30)         & tv17 (30)           & tv18  (30)          & tv19 (30)   &tv20 (20) &tv21 (20) &tv22 (30) &tv23 (20)  &Mean     \\
                          \hline
 Reformulation agent ($f_Q$) & 0.284&0.337&0.198& 0.254& 0.314&0.390&0.239 &0.311 &0.291  \\
 + Memory bank and decision reasoning ($\hat{f}_Q$)  &\textbf{0.292}&\textbf{0.357}&0\textbf{.211}&\textbf{0.280} &\textbf{0.370} & \textbf{0.418} &\textbf{0.257 }&\textbf{0.335} &\textbf{0.315}\\
\bottomrule
\end{tabular}
}
    \vspace{-0.15in}

\end{table*}

\begin{figure}
  \centering
        \includegraphics[width=0.95\linewidth]{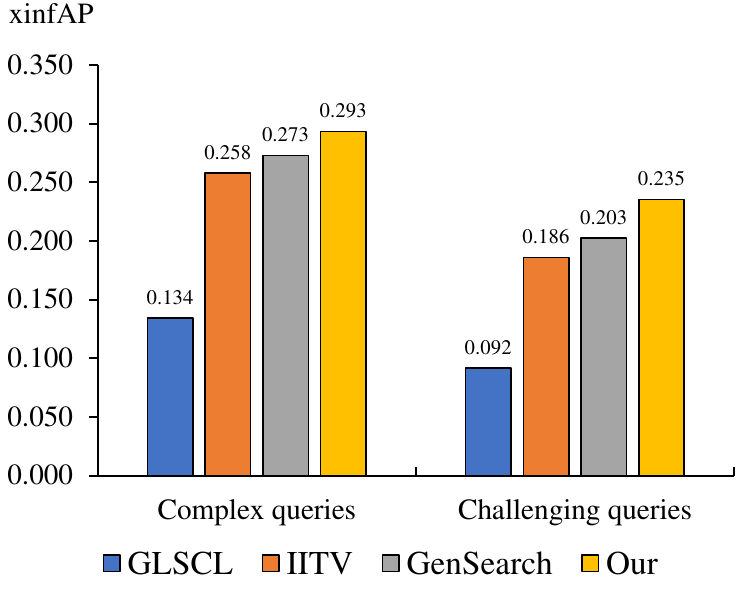}
      \caption{Performance comparison on complex and more challenging queries that require multi-step reasoning. }
\label{fig:barchart_complex_query_comparison}
    \vspace{-0.15in}

\end{figure}

\begin{figure}

  \centering
        \includegraphics[width=1\linewidth]{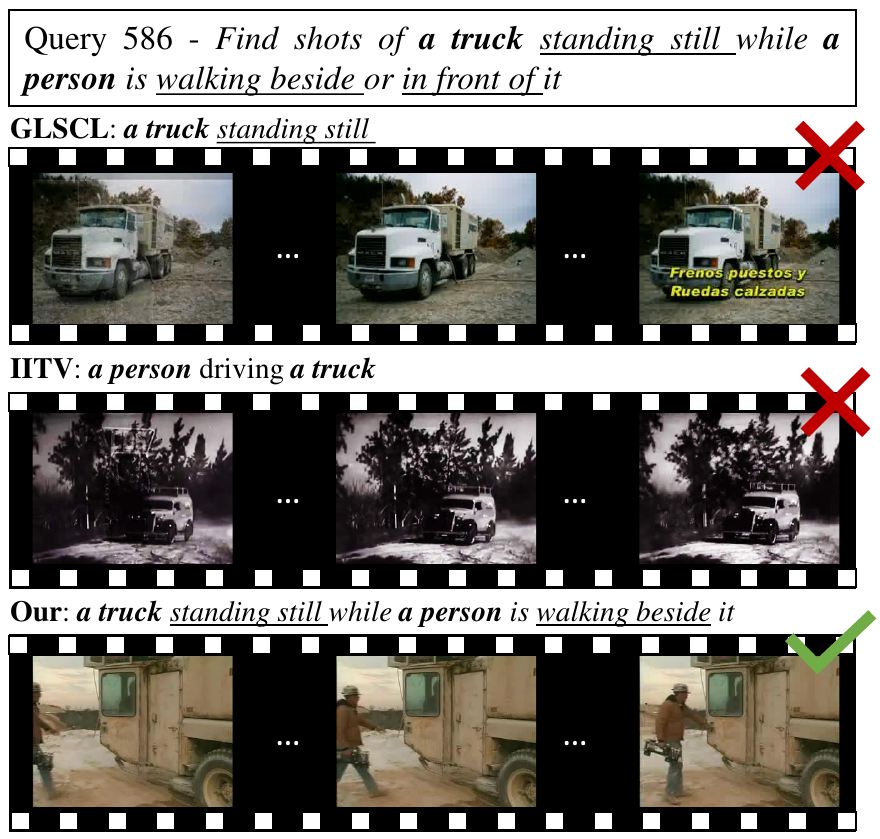}
      \caption{Rank-1 videos retrieved by GLSCL, IITV and our model, respectively. With contextual temporal reasoning, our model is able to rank correct result the highest. }
\label{fig:complex_query_visualization}
    \vspace{-0.25in}

\end{figure}

We further analyze performance across different query types, focusing on complex queries to validate the merit of contextual reasoning. Specifically, we asked GPT-4 to identify the complex queries from 210 test queries.
Complex queries are defined as those that require temporal reasoning and cannot be reliably addressed using static visual information alone.
Figure~\ref{fig:barchart_complex_query_comparison} presents performance comparisons on this subset. For the complex queries (114 out of 210), our multi-agent retrieval framework outperforms GLSCL~\cite{zhangTextVideoRetrievalGLSCL2025}, IITV~\cite{improvedITV}, and GenSearch~\cite{wu2025genSearch} on 100\%, 86.6\%, and 80.7\% of the queries, respectively. 
For example, for query 567 — \textit{Find shots of people performing or dancing outdoors at nighttime}, the performance gains over GLSCL, IITV, and GenSearch are 940\%, 66\%, and 74\%, respectively.
Additionally, we asked GPT-4 to identify more challenging queries involving temporal relationships, logical constraints, causal reasoning, ambiguity, or multi-step understanding. A total of 44 such queries were identified, and the corresponding performance comparison is shown in Figure~\ref{fig:barchart_complex_query_comparison}. On this more challenging subset, our method outperforms GLSCL, IITV, and GenSearch on 93.2\%, 88.6\%, and 77.3\% of the queries, respectively.
For example, for query 586 — \textit{Find shots of a truck standing still while a person is walking beside or in front of it},  the performance gains over GLSCL, IITV, and GenSearch are 662\%, 82\%, and 175\%, respectively. 
Figure~\ref{fig:complex_query_visualization} visualizes the top-1 retrieved video for this query by GLSCL, IITV, and our method. GLSCL ranks a video of a stationary truck without any interaction as the most relevant result. IITV retrieves a video showing a person driving a truck, which partially matches the query but lacks the required spatial dynamics. In contrast, our method, empowered by contextual temporal reasoning, successfully retrieves a video where a person is walking beside a stationary truck, precisely fulfilling the query condition.

\begin{table*}[t]
\caption{The performance boost over the proposed agents.} 
\label{tab:agent_comparison}
\centering
\resizebox{\linewidth}{!}{
\begin{tabular}{l|ccc|ccc|cc|c}
    \toprule
Datasets                     & \multicolumn{3}{c|}{IACC.3}               & \multicolumn{3}{c|}{V3C1}   & \multicolumn{2}{c|}{V3C2}   \\
\hline
Query sets                         & tv16 (30)         & tv17 (30)           & tv18  (30)          & tv19 (30)   &tv20 (20) &tv21 (20) &tv22 (30) &tv23 (20)  &Mean     \\
                          \hline
Retrieval Agent & 0.271 &0.343 &0.164 &0.243 &0.347 &0.357 &0.219 &0.294 &0.281\\
+ Reasoning Agent  &0.293 & 0.357 & 0.200 &\textbf{0.292 }&\textbf{0.381} &0.416 &\textbf{0.273} & 0.340 &0.318\\
+ Reformulation Agent  &0.292&0.357&0.211&0.280 &0.370 & 0.418 &0.257 &0.335 &0.315 \\
+ Orchestration Agent  &\textbf{0.298} &\textbf{0.371} & \textbf{0.219}&0.283&0.379& \textbf{0.420}& 0.271&\textbf{0.342}& \textbf{0.323}\\

\bottomrule
\end{tabular}
}

\end{table*}

\begin{table*}[t]
\caption{The Sensitivity of the hyperparameter $T$ and $k$.} 
\label{tab:sensitivity}
\centering
\resizebox{\linewidth}{!}{
\begin{tabular}{l|ccc|ccc|cc|c}
    \toprule
Datasets                     & \multicolumn{3}{c|}{IACC.3}               & \multicolumn{3}{c|}{V3C1}   & \multicolumn{2}{c|}{V3C2}   \\
\hline
Query sets                         & tv16 (30)         & tv17 (30)           & tv18  (30)          & tv19 (30)   &tv20 (20) &tv21 (20) &tv22 (30) &tv23 (20)  &Mean     \\
                          \hline


$T=60$ and $k=50$&0.298 &\textbf{0.371}&0.219 &0.283 &0.379 &0.420&0.271 &0.342 &0.323\\

$T=30$ and $k=100$&\textbf{0.299} &0.364 & 0.211&0.291&\textbf{0.383}& 0.420 &0.273 &0.338 &0.322\\
$T=50$ and $k=100$ &0.297&0.367&\textbf{0.223}&\textbf{0.293}&0.382&\textbf{0.423}&\textbf{0.275}&\textbf{0.343} &0.325\\

\bottomrule
\end{tabular}
}
    \vspace{-0.15in}

\end{table*}

\subsection{Ablation studies}

\subsubsection{Effectiveness of the communication mechanism}

In this section, we evaluate the effectiveness of the proposed memory components $M$ through ablation studies.
Table~\ref{tab:query_reformulation_comparison} compares two versions of the query reformulation agent: the baseline version defined in Eq.(\ref{eq:reformulation_agent_init}), and the enhanced version defined in Eq.(\ref{eq:reformulation_agent_new}). 
The results show that incorporating the memory bank and decision reasoning consistently improves performance across all datasets and query sets. On average, the enhanced reformulation agent achieves a mean xinfAP of 0.315, outperforming the baseline (0.291).

\subsubsection{Effectiveness of the Proposed Agents}

In this section, we evaluate the individual contributions of each proposed agent within our multi-agent retrieval framework. We begin with the retrieval agent as the baseline and incrementally add other agents to assess their impact on overall performance.
First, we add the contextual reasoning agent. In this setting, we adopt a greedy strategy in which only exploitation is performed in each iteration. Next, we incorporate the query reformulation agent alongside the retrieval and reasoning agents. Here, we also use a greedy reformulation strategy that modifies the query in every iteration. Finally, we include the orchestration agent, which dynamically determines whether to continue exploitation of current rank list or trigger exploration for a new query, yielding the complete multi-agent retrieval system.
Table~\ref{tab:agent_comparison} compares performance across eight TRECVid AVS query sets. The addition of the contextual reasoning agent yields significant improvement over the retrieval-only baseline across all datasets. Specifically, it outperforms the baseline on 176 out of 210 queries, and over 50\% of these improved queries are categorized as complex by GPT-4. The average improvement rate across all queries is 43.8\%.
The inclusion of the query reformulation agent improves performance on 165 out of 210 queries when compared to the retrieval-only baseline, with 95 of these being complex queries, particularly in certain query sets (e.g., tv18 and tv21). However, it leads to performance drops when compared with the combination of the retrieval agent and the contextual reasoning agent. We observe that most of the queries with performance declines are relatively easy ones where the retrieval agent already performs well. In these cases, query modification may overly constrain the search space, reducing the likelihood of retrieving relevant results.

\begin{figure}[t]

  \centering
        \includegraphics[width=1\linewidth]{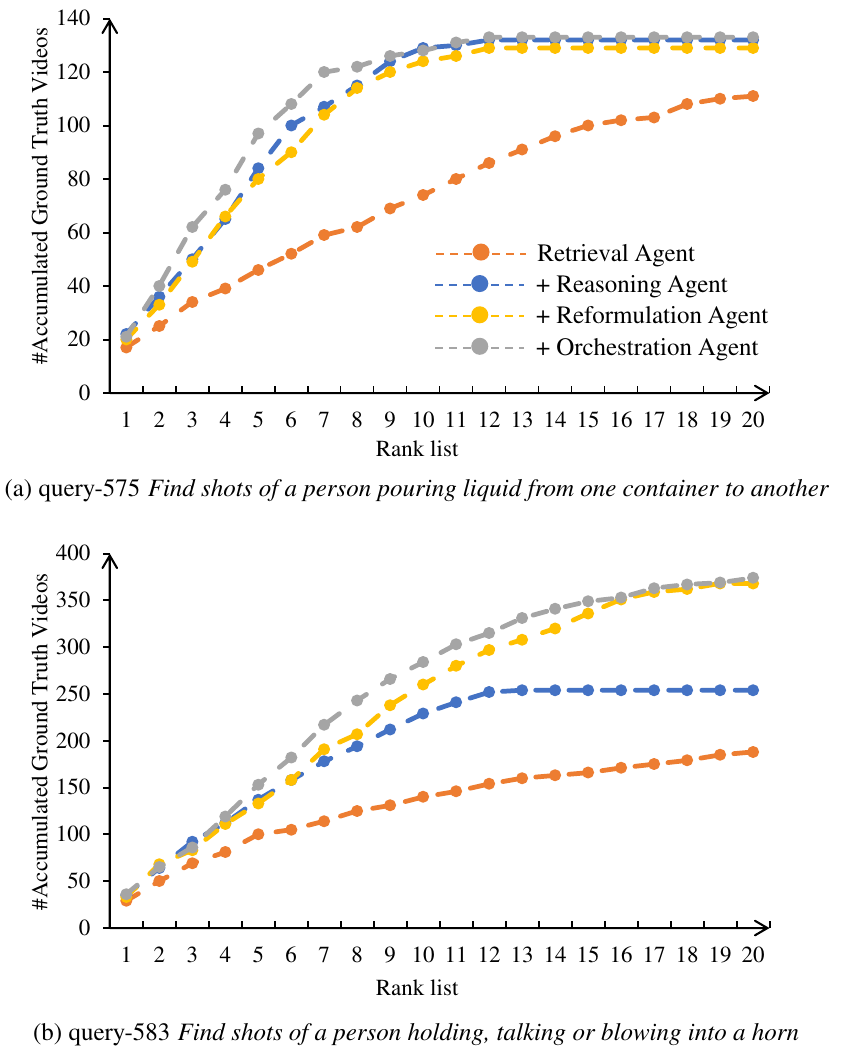}
      \caption{Performance of the adaptive agentic workflow (with Orchestration Agent) versus fixed greedy strategies, measured by the number of accumulated ground truth videos over the ranked list. }
\label{fig:accumulated_GT}
    \vspace{-0.2in}

\end{figure}
When the orchestration agent is included to determine the contextually optimal execution plan at each iteration, significance tests show that the improvements are statistically significant compared to the two previously used greedy strategies across all 210 queries.
Figures~\ref{fig:accumulated_GT}(a) and~(b) further compare the speed and number of accumulated ground truth videos for one complex and one ambiguous query. The rank list is divided into 20 bins, and each bin represent 50 candidate videos. The retrieval-only agent serves as the baseline. The orchestration agent can take advantages of either reasoning agent or the reformulation agent to retrieve ground truth videos more effectively and efficiently.

\subsubsection{Sensitivity to Hyperparameters}

We conduct a sensitivity analysis to examine how different values of the maximum iterations $T$ and the examination window size $k$ affect retrieval performance. Table~\ref{tab:sensitivity} reports results under three hyperparameter configurations.
When increasing the window size from $k = 50$ to $k = 100$ and reducing the number of iterations from $T = 60$ to $T = 30$, while keeping the total number of examined videos ($k \times T$) approximately constant to ensure fair comparison, we observe a slight decrease in average performance (from 0.323 to 0.322). This suggests that more frequent updates with smaller windows may be marginally more effective, potentially due to finer-grained feedback and more adaptive exploration.
Interestingly, when both $T$ are increased to 50, the mean performance slightly improves to 0.325, indicating a trade-off between search depth and the frequency of action adjustment. 
Overall, our framework demonstrates relatively robustness to variations in $T$ and $k$.


\section{Conclusion and Future Work}

We propose a dynamic multi-agent retrieval framework to address the contextual and temporal reasoning challenges posed by complex and ambiguous queries in text-to-video retrieval. Extensive experiments demonstrate that the adaptive agentic workflow significantly outperforms existing state-of-the-art methods and greedy strategies across both standard and complex query sets.
Future work may explore reinforcement learning for agent coordination and user-in-the-loop feedback mechanisms.



{
    \small
    \bibliographystyle{ieeenat_fullname}
    \bibliography{main}
}


\appendix
\clearpage


\section{Multi-agent Retrieval Prompts}
In the following, we provide the prompts (i.e., $p_{\text{eval}}$, $p_{\text{refine}}$, $p_{\text{action}}$, and $\hat{p}_{\text{refine}}$) we used in the multi-agent retrieval framework. These prompts guide the behavior of different agents in performing video evaluation, reformulation, decision-making, and refinement. Note that we have both $p_{\text{eval}}$ and $p_{\text{evalReasoning}}$ to balance efficiency and interpretability.

\begin{tcolorbox}[boxrule=0pt, frame empty]
$p_{\text{action}} =$ 

\textbf{Observation}:

Query: \{query\}

TopK eval summary: \{eval\_summary\}

\textbf{Question}: Is the current rank list satisfactory?

1) Answer with \texttt{"exploit"} if the current rank list is good and it is worth exploring further.

2) Answer with \texttt{"explore"} if the current rank list is not good and the query should be refined to refresh the results.

\textbf{Answer (in JSON format)}:

\{
  \texttt{"action"}: \texttt{"exploit"} or \texttt{"explore"}, \\
  \texttt{"reasoning"}: \texttt{"brief reasoning of the action"}
\}
\end{tcolorbox}

\begin{tcolorbox}[boxrule=0pt, frame empty]
$p_{\text{eval}} =$ 

You are a helpful assistant who can judge whether a video is relevant to a given query.

\textbf{Query:} \{query\} \\
\textbf{Video path:} \{Video\_path\}

Please judge whether the video is relevant to the query.

\textbf{Answer with only one word:} \texttt{"matched"} or \texttt{"unmatched"}.
\end{tcolorbox}

\begin{tcolorbox}[boxrule=0pt, frame empty]
$p_{\text{evalReasoning}} =$ 

You are a helpful assistant who can judge whether a video is relevant to a given query.

\textbf{Query:} \{query\} \\
\textbf{Video path:} \{Video\_path\}

Please judge whether the video is relevant to the query.

\textbf{Answer (in JSON format)}:

\{
  \texttt{"Evaluation"}: \texttt{"matched"} or \texttt{"unmatched"}, \\
  \texttt{"reasoning"}: \texttt{"brief reasoning of the evaluation"}
\}
\end{tcolorbox}

\begin{tcolorbox}[boxrule=0pt, frame empty]
$p_{\text{refine}} =$ 

\textbf{Instruction:} You are a helpful query reformulator. Reformulate the query to increase the chance of matching the target videos.

\textbf{INPUT:}  
The user query is: \{original\_query\}.  
The current search query is: \{query\}.

\textbf{OUTPUT:}  
Output the reformulated query (no more than 30 words) in the following format:

\texttt{<reformulate>}  
\texttt{[Reformulated query here]}  
\texttt{</reformulate>}
\end{tcolorbox}

\begin{tcolorbox}[boxrule=0pt, frame empty]
$\hat{p}_{\text{refine}} =$ 

You are a helpful query reformulator. Reformulate the query to increase the chance of matching the target videos.

You will be provided with some queries and their precision as obtained from the search model.  
Additionally, the reason for needing to reformulate the query is also provided.  
You can refer to this information when reformulating.

\textbf{Suggestion:}  

- If the precision of the query is high, you are advised to make only small changes (e.g., make it more specific) or keep the original query. 

- If the precision is low, imagine relevant scenarios and describe them in your reformulated query  while preserving the original meaning of the user query.

\textbf{Important Instructions:}  

- \textbf{Do not use negation words} (e.g., "not", "no", "without") — the search model performs poorly with negation.  

- Reformulate the query to maximize \textbf{semantic match} with target videos.  

- \textbf{Preserve the original meaning} of the user query.  

- The reformulated query should be \textbf{less than 30 words}.

\textbf{INPUT:}  

The query and its precision across steps: \{ \texttt{self.query\_performance\_memory\_bank} \}  

Reason for reformulation: \{ \texttt{action\_decision\_reasoning} \}  

The user query: \{ \texttt{original\_query} \}  

The current search query: \{ \texttt{query} \}

\textbf{Output Format:}

\texttt{<think>}  
Your reasoning process: what changed, why, and how it improves matching.  
\texttt{</think>}  

\texttt{<reformulate>}  
[Reformulated query here]  
\texttt{</reformulate>}
\end{tcolorbox}

\begin{table*}[t]
\caption{The performance improvement of using another cross-modal retrieval (InternVid \cite{wang2023internvid}).} 
\label{tab:viclip}
\centering
\resizebox{\linewidth}{!}{
\begin{tabular}{l|ccc|ccc|cc|c}
    \toprule
Datasets                     & \multicolumn{3}{c|}{IACC.3}               & \multicolumn{3}{c|}{V3C1}   & \multicolumn{2}{c|}{V3C2}   \\
\hline
Query sets                         & tv16 (30)         & tv17 (30)           & tv18  (30)          & tv19 (30)   &tv20 (20) &tv21 (20) &tv22 (30) &tv23 (20)  &Mean     \\
                          \hline
InternVid \cite{wang2023internvid} & 0.185 & 0.216 &0.104 &0.122 &0.199 &0.190 &0.125 &0.103 & 0.155\\
 Multi-agent Retrieval &\textbf{0.221} &\textbf{0.272} &\textbf{0.181} &\textbf{0.172}&\textbf{0.256}&\textbf{0.260} &\textbf{0.197}&\textbf{0.173}& \textbf{0.217}\\
\bottomrule
\end{tabular}
}

\end{table*}

\begin{figure*}
    \centering
        \includegraphics[width=1\linewidth]{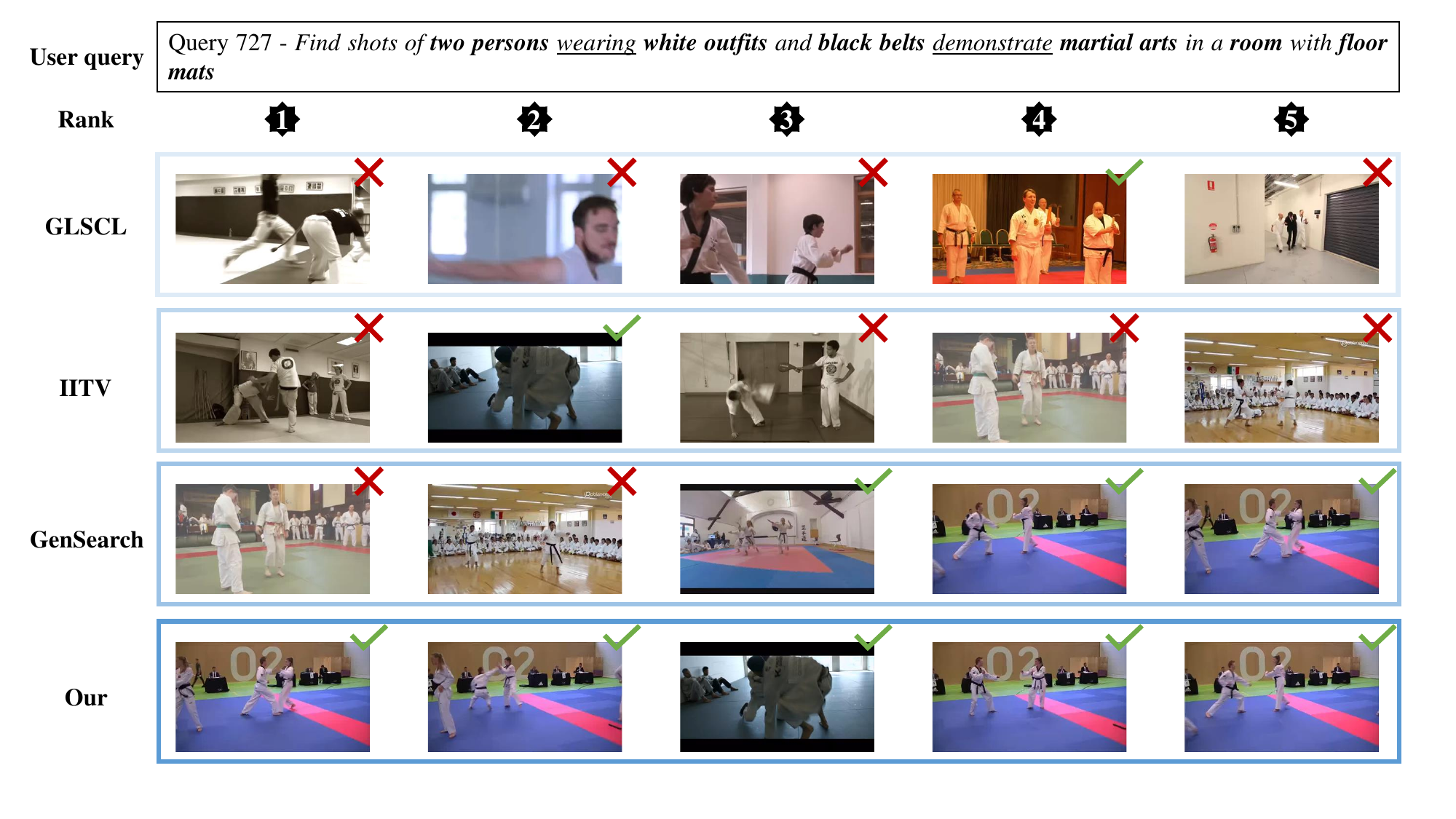}
    \caption{Top-5 ranked video comparison with recent methods on a complex query that need multi-round reasoning, GLSCL \cite{zhangTextVideoRetrievalGLSCL2025}, IITV \cite{improvedITV}, and GenSearch \cite{wu2025genSearch}. }
    \label{fig:top5comparison}
\end{figure*}

\section{Experimental Result with Other  Retrieval Agent }
We also validate our proposed framework using another cross-modal retrieval model, InternVid \cite{wang2023internvid}. The results demonstrate that the retrieval performance improvements achieved by our multi-agent retrieval framework are consistent across different retrieval backbones. Specicially, the results show that our multi-agent framework improves the mean retrieval score from 0.155 to 0.217, a relative gain of  40\%.


\section{Case Studies}

We provide more case studies in this section to verify the merit of our proposed multi-agent retrieval framework. Figure \ref{fig:top5comparison} visulizes the top-5 videos retrieved by recent models, including GLSCL \cite{zhangTextVideoRetrievalGLSCL2025}, IITV \cite{improvedITV}, and GenSearch \cite{wu2025genSearch}, for the query-727 \textit{Find shots of two persons wearing white outfits and black belts demonstrate martial arts in a room with floor mats}. Unlike other models that focus on partial or superficial matches, e.g., retrieving scenes with people in white outfits but lacking the correct number of persons, action context, or floor mats (as seen in GLSCL and IITV), our proposed method accurately attends to all critical components of the query, including the number of persons, attire, action semantics, and environmental context, demonstrating better multi-round reasoning and cross-modal alignment.

\begin{figure*}
    \centering
        \includegraphics[width=1\linewidth]{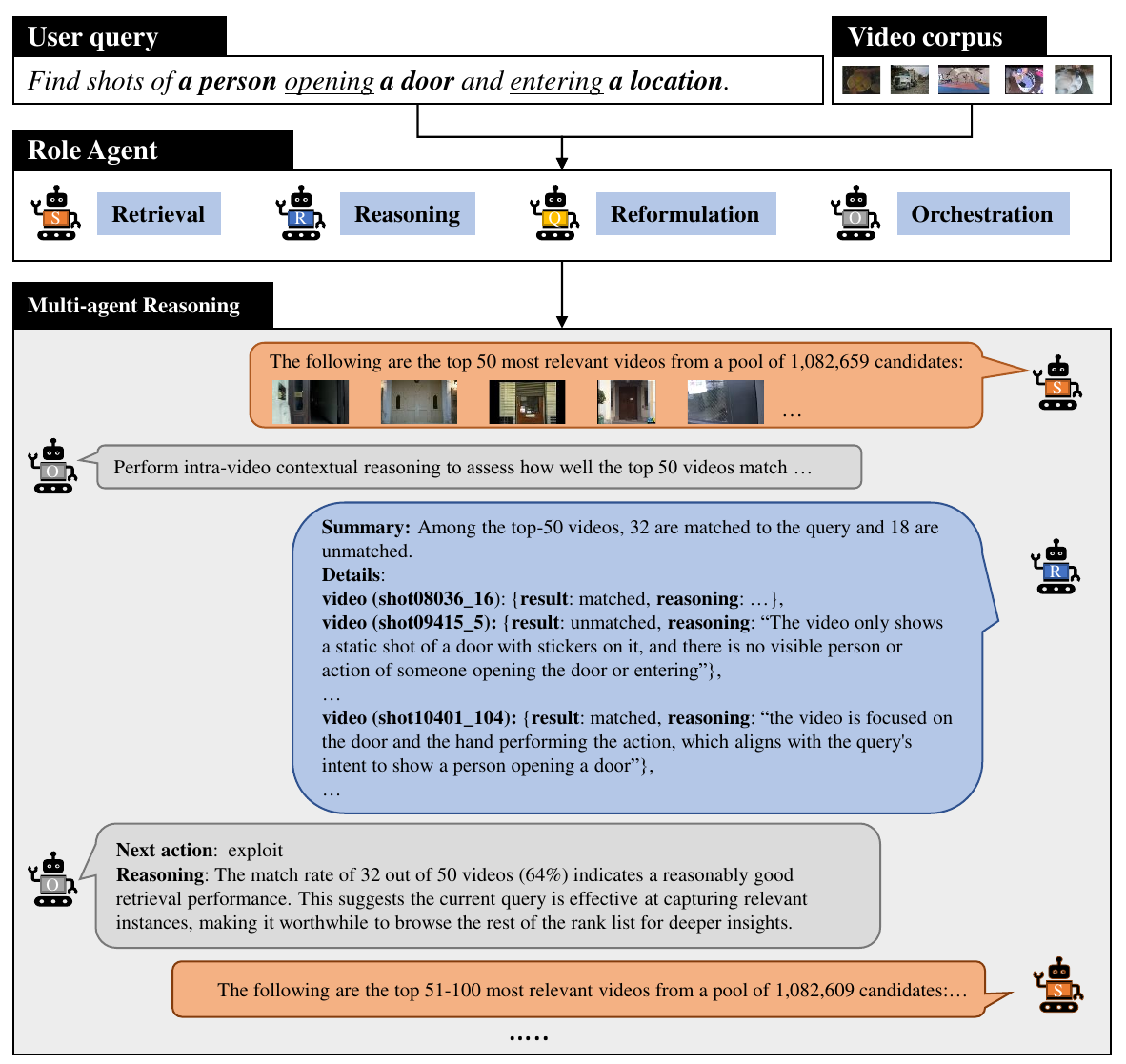}
    \caption{Example of multi-agent retrieval reasoning for the query \textit{Find shots of a person opening a door and entering a location}.}
    \label{fig:reasoning_process1}
\end{figure*}

\begin{figure*}
    \centering
        \includegraphics[width=1\linewidth]{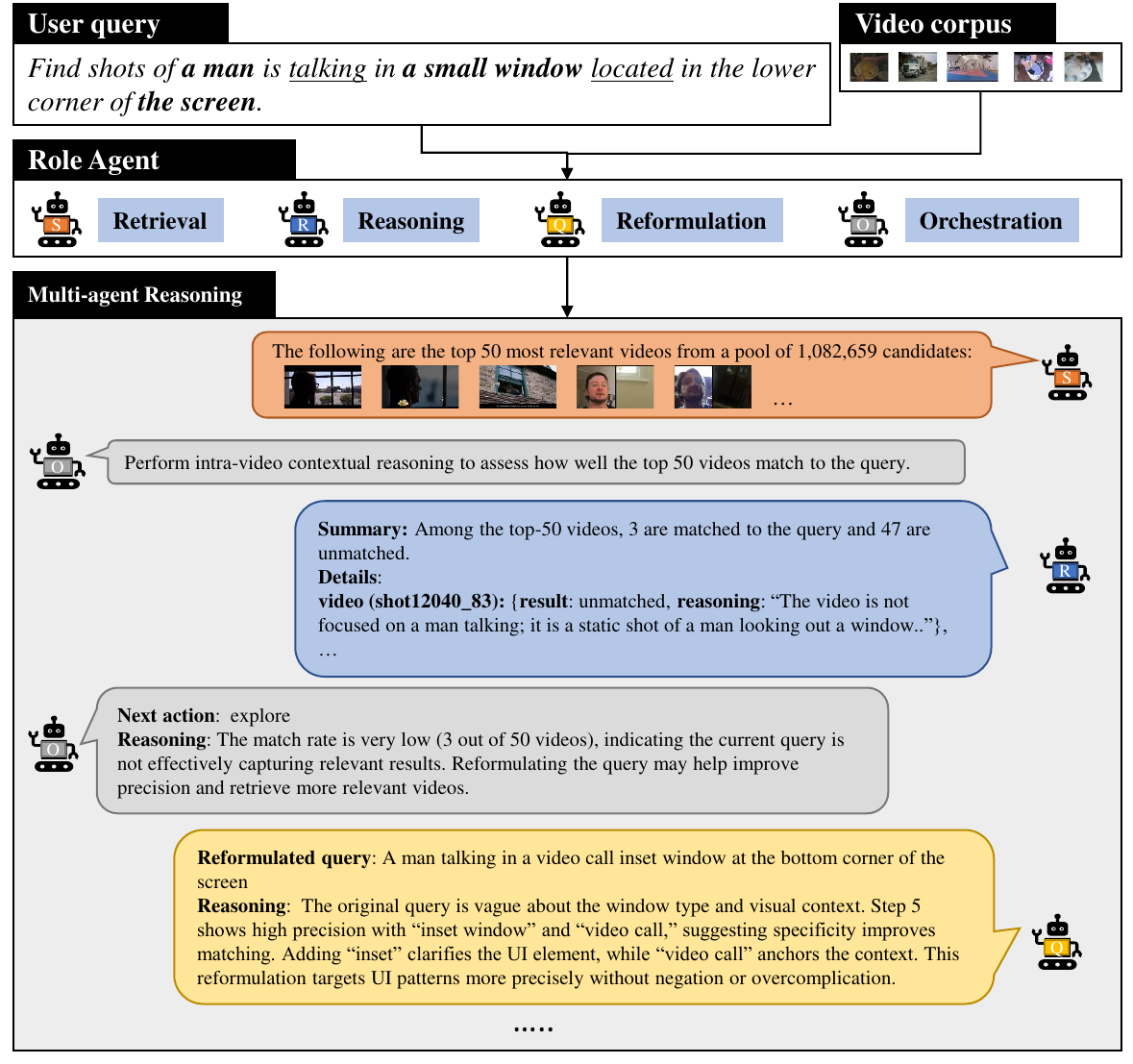}
    \caption{Example of multi-agent retrieval reasoning for the query \textit{Find shots of a man is talking in a small window located in the lower corner of the screen}.}
    \label{fig:reasoning_process2}
\end{figure*}

Figures~\ref{fig:reasoning_process1} and~\ref{fig:reasoning_process2} illustrate how our multi-agent system collaboratively performs reasoning and decision-making to address complex video retrieval queries. Each agent contributes specialized capabilities, i.e., retrieval, reasoning, reformulation, and orchestration, working together to interpret user intent and refine results dynamically.
Figure~\ref{fig:reasoning_process1} shows the reasoning process for the query \textit{Find shots of a person opening a door and entering a location}. This query demands both semantic decomposition (i.e., identifying a person, a door, and a location) and temporal reasoning (i.e., verifying that opening precedes entering). The retrieval agent initially finds 50 candidate videos, but the reasoning agent determines that only 32 (64\%) truly satisfy the temporal and semantic constraints. For example, static scenes such as shot09415\_5 are rejected, while dynamic sequences like shot10401\_104 are accepted for accurately depicting the intended action sequence. The orchestration agent, informed by this evaluation, decides to exploit the current query and continue retrieval without reformulation. This case illustrates how multi-agent collaboration enables fine-grained reasoning over temporal dynamics and supports adaptive strategy selection.
Figure~\ref{fig:reasoning_process2} demonstrates the system's collaborative reasoning on a visually grounded and spatially specific query: \textit{Find shots of a man is talking in a small window located in the lower corner of the screen}. Unlike the temporal reasoning in the previous case, this query challenges the system’s ability to interpret UI layout, spatial positioning, and scene composition, which are often ambiguous in natural language. Although the retrieval agent surfaces 50 candidates, the reasoning agent finds that only 3 (6\%) match the intended visual structure. Recognizing the low match rate, the orchestration agent triggers the reformulation agent, which generates a more precise query “A man talking in a video call inset window at the bottom corner of the screen.” This reformulation introduces UI-specific terms like “inset window” and “video call,” improving alignment with visual patterns in the corpus. This example highlights how multi-agent collaboration enables visual disambiguation and query refinement, leading to more accurate and efficient retrieval in complex visual environments.

\end{document}